\pdfoutput=1

\documentclass[11pt]{article}

\usepackage[preprint]{acl}

\usepackage{times}
\usepackage{latexsym}

\usepackage[T1]{fontenc}

\usepackage[utf8]{inputenc}

\usepackage{microtype}

\usepackage{inconsolata}

\usepackage{soul}
\usepackage{graphicx}
\usepackage{booktabs}
\usepackage{float}
\usepackage{overpic}
\usepackage{float}  
\usepackage{subfloat}
\usepackage{stfloats} 
\usepackage[skip=0.5\baselineskip]{caption}
\usepackage{bm}
\usepackage{amsmath}
\usepackage{booktabs}
\usepackage{multirow}
\usepackage{multicol}
\usepackage{enumitem}
\usepackage{subcaption}
\usepackage{threeparttable}
\usepackage{xspace}
\usepackage{color}
\usepackage[normalem]{ulem}
\usepackage{algorithmicx,algorithm}
\usepackage{algpseudocode}
\usepackage{algorithm,algpseudocode}
\usepackage{marvosym}   
\usepackage{colortbl}   
\usepackage{tcolorbox}  
\tcbuselibrary{breakable}   
\usepackage{wasysym}    
\usepackage[figuresright]{rotating} 

\usepackage{xspace}
\newcommand{\etal}{\emph{et al.}\xspace}
\newcommand{\eg}{\emph{e.g.,}\xspace}
\newcommand{\ie}{\emph{i.e.,}\xspace}

\newcommand{\name}{ReaEmb\xspace}
\makeatletter
\newcommand\blfootnote[1]{%
  \begingroup
  \renewcommand\thefootnote{}%
  \long\def\@makefntext##1{\parindent 1em\noindent ##1}%
  \footnotetext{#1}%
  \endgroup
}
\makeatother

\title{Harmonizing Semantic and Collaborative in LLMs: \\ Reasoning-based Embedding Generator for Sequential Recommendation}

\author{
  \textbf{Qidong Liu\textsuperscript{*}, Mingyao Huang\textsuperscript{*}, Moranxin Wang, Wenxuan Yang, Haiping Zhu} \\
  Xi'an Jiaotong University, Xi'an, China \\
  \texttt{\{liuqidong,zhuhaiping\}@xjtu.edu.cn} \\
  \texttt{\{2214312088,wangmo,yangwenxuan\}@stu.xjtu.edu.cn} \\
}

\begin{document}

\maketitle
\blfootnote{\textsuperscript{*}Both authors contributed equally. }
\setcounter{footnote}{0}

\begin{abstract}
Sequential Recommender Systems (SRS) predict the next item of interest based on users' interaction histories and have been widely deployed, but hindered by long-tail problem.
Large Language Models (LLMs), with strong semantic understanding and reasoning capabilities, offer a promising way to enrich item semantics and have recently been used as embedding generators.
However, two fundamental gaps remain.
First, current LLM-based embedding methods fail to exploit the model's inner reasoning capacity.
Second, existing methods often inject collaborative signals implicitly via supervised fine-tuning, lacking explicit guidance for collaborative embedding alignment.
In this paper, we introduce \textbf{\name}, a novel framework that resolves both issues via a \emph{Latent Reasoning-enhanced Contrastive Learning} (LRCL) stage and a \emph{Collaborative Reward Reinforcement Learning} (CRRL) stage. LRCL exploits the LLM’s inner reasoning capacity through a two-pass forward process with an additional attention module.
CRRL subsequently explicitly injects collaborative signals into the LLM via a tailored reinforcement learning. 
Extensive experiments on three real-world datasets demonstrate superior effectiveness of \name across multiple SRS models.
To ease reproducibility, we release the code online\footnote{https://github.com/mingyao-huang/ReaEmb.git}.
\end{abstract}

\section{Introduction}

Sequential Recommender Systems (SRS) aim to predict the next item a user is most likely to interact with based on the historical interaction sequence~\cite{fang2020deep, wang2019sequential}, and have been widely applied in e-commerce~\cite{wang2020time}, short-video platforms~\cite{pan2023understanding}, and online education~\cite{zhang2019hierarchical}. Deep SRS models, \eg SASRec~\cite{kang2018self} and BERT4Rec~\cite{sun2019bert4rec}, have shown strong capability in capturing users' dynamic preferences. 
However, these models still suffer from the long-tail problem~\cite{wang2019sequential}, where a few popular items dominate interactions while most items receive very few.
Such a data distribution makes current SRS difficult to learn expressive collaborative representation for long-tail items, leading to degraded accuracy.

To alleviate long-tail problem, incorporating item semantics has been shown effective~\cite{liu2025large, guo2020survey}. Large Language Models (LLMs) possess broad semantic knowledge beyond item popularity, making them compelling generators of item embeddings for SRS~\cite{liu2025llmemb, he2025llm2rec, lee2025llm}. 
Following this way, recent works proposed prompting LLMs to output embeddings for SRS models. Despite this promise, two critical gaps persist in existing LLMs-based embedding approaches.

\begin{figure}[t]
    \centering
    \makebox[\columnwidth][c]{%
        \includegraphics[width=1.1\columnwidth]{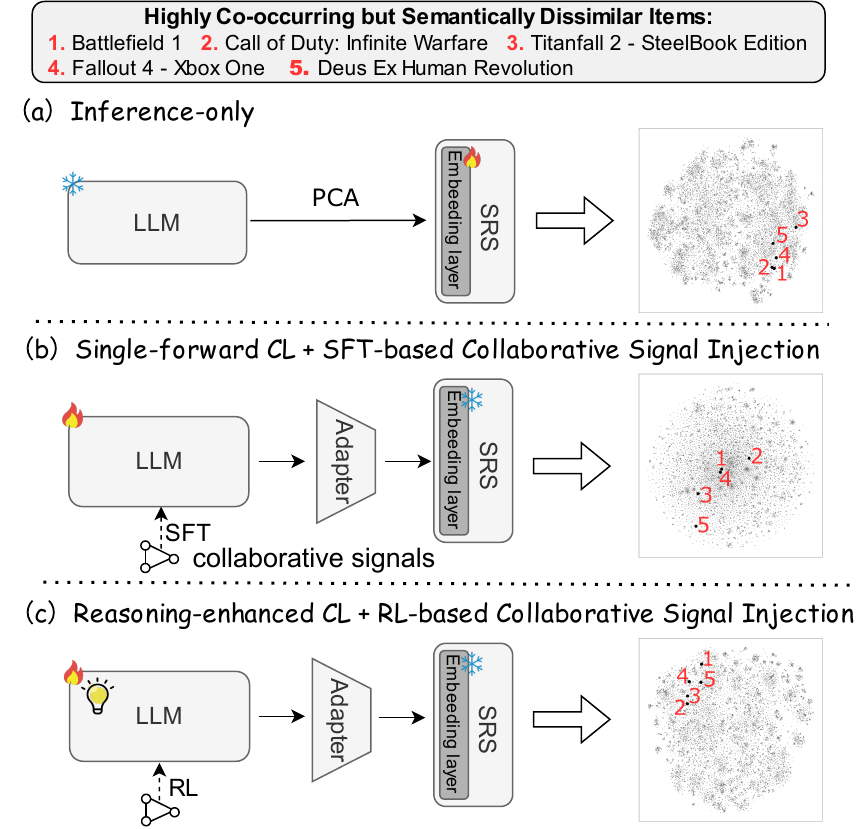}
 }
    \caption{Comparison of three paradigms for adapting LLMs into item embedding generators.}
    \label{fig:case_study}
\vspace{-16pt} 
\end{figure}

\begin{itemize}[leftmargin=*, topsep=0pt, itemsep=0pt, parsep=0pt, partopsep=0pt]
    \item \textbf{Insufficient Semantic Understanding.} Current methods obtain item representations through a \emph{single} LLMs forward pass~\cite{liu2025llmemb,he2025llm2rec}, treating the LLMs mainly as a text encoder. This paradigm underutilizes the model's intrinsic reasoning ability~\cite{jaech2024openai}, limiting its capacity to uncover fine-grained item attributes and implicit semantic relations.

    \item \textbf{Implicit Collaborative Signal Injection.} Collaborative signals, which reflect item co-occurrence patterns, are essential for recommendation~\cite{he2020lightgcn}. However, existing LLM-based embedding methods still incorporate them indirectly. As shown in Figure~\ref{fig:case_study}(a), the \textbf{Inference-only Paradigm}~\cite{hu2024enhancing} directly learns collaborative signals into LLMs embeddings by co-training with SRS models. However, as visualized in the t-SNE plot, fine-tuning LLMs embeddings may lead to the loss of semantic information, causing highly co-occurring items to cluster together. 
    The \textbf{SFT-based Injection} paradigm in Figure~\ref{fig:case_study}(b) adopts supervised fine-tuning on next token prediction tasks to implicitly inject collaborative signals, while incorporating a trainable adapter to prevent semantic loss~\cite{he2025llm2rec}. 
    Nevertheless, LLMs embeddings are still dominated by semantics, leaving co-occurring items far apart in latent space. Such a gap between semantic and collaborative is hard to narrow only by an adapter. 
\end{itemize}

In terms of the first issue, LLMs reasoning is a promising solution.
However, most reasoning methods require explicit linguistic chains to inspire LLMs, which is difficult to evaluate, especially for embedding task. Such a \textbf{Linguistic Gap} makes existing reasoning methods unsuitable for SRS embedding generation.
Inspired by recent advances in latent reasoning~\cite{zhang2025reinforced}, we introduce an implicit reasoning mechanism that performs reasoning in the hidden representation space to produce semantic-enriched item embeddings.
As for the collaborative signal injection challenge, reinforcement learning provides a potential solution. 
However, existing RL designs for LLMs mainly target text retrieval, matching, or generation tasks, where rewards are derived from textual relevance or answer quality, denoted as \textbf{Mismatch Reward}. Such text-oriented rewards cannot directly guide the geometric optimization of item embeddings. To bridge this gap, we design a co-occurrence-based reward that encourages frequently co-occurring items to be closer in the embedding space. 

Based on the design above, we propose \textbf{\name}, a novel two-stage LLMs training framework for high-quality item representation. 
As shown in the t-SNE visualization in Figure~\ref{fig:case_study}(c), \name aims to generate embeddings that jointly reflect collaborative relations and preserve semantic distinguishability.
In the first stage, named \emph{Latent Reasoning-enhanced Contrastive Learning} (LRCL), we introduce an implicit reasoning module~\cite{zhang2025reinforced} that simulates latent reasoning with several reasoning tokens and an additional attention layer. 
In the second stage, called \emph{Collaborative Reward Reinforcement Learning} (CRRL), we apply a GRPO-based~\cite{shao2024deepseekmath} reinforcement learning procedure that rewards highly co-occurring items for being closer in the embedding space, thereby explicitly injecting collaborative signals into the LLMs. 
The contributions of this paper are as follows:
\begin{itemize}[leftmargin=*, topsep=0pt, itemsep=0pt, parsep=0pt, partopsep=0pt]
    \item We propose \name, a two-stage training framework that transforms LLMs into high-quality embedding generators.
    \item To address insufficient semantic understanding, we introduce an implicit reasoning module to inspire LLMs before embedding generation. 
To explicitly inject collaborative signals, we introduce a co-occurrence-based reward for RL.
    \item We conduct extensive experiments on three real-world datasets across multiple SRS backbones, demonstrating the effectiveness of \name.
\end{itemize}

\section{Preliminary}
\textbf{Sequential Recommendation} \\
Let $\mathcal{U}$ and $\mathcal{V}$ denote the sets of users and items, respectively. For each user $u \in \mathcal{U}$, the chronologically ordered interaction sequence is $S_u = [v_1^{(u)}, v_2^{(u)}, \ldots, v_{|S_u|}^{(u)}]$ where $v_t^{(u)} \in \mathcal{V}$. For simplicity, we omit the user-specific superscript $(u)$ in subsequent notations. The goal of sequential recommendation is to predict the next item the user will interact with:
\begin{equation}
    \hat{v} = \arg\max_{v \in \mathcal{V}} P\!\left(v_{|S_u|+1} = v \mid S_u\right).
\end{equation}
\textbf{Embedding Generator for SRS} \\
Most sequential recommendation models follow a general \textit{Embedding-Sequence} framework. The embedding function first transforms each item $v_i \in \mathcal{V}$ into a dense vector $\mathbf{e}_{i}=\mathrm{Emb}(v_i)$, and correspondingly converts the user sequence $S_u$ into an embedding sequence $\tilde{S}_u$. The resulting embedding sequence is then modeled by an SRS backbone $\mathrm{Seq}(\cdot)$ to extract the user's dynamic preference for next-item prediction.
In conventional SRS methods, $\mathrm{Emb}(\cdot)$ is often implemented as a randomly initialized ID embedding table and learned from interaction data. In LLMs-based embedding methods~\cite{liu2025llmemb}, each item $v_i$ is associated with a textual description $T_{i}$, including attributes such as title, category and brand. The embedding function can then be instantiated by encoding $T_{i}$ with an LLMs: $\mathbf{e}_{i}={\mathrm{LLM}}(T_{i})$. In this paper, we focus on fine-tuning the LLMs to generate high-quality item embeddings that can better support downstream sequential recommendation.
\section{Method}
\label{sec:method}

\subsection{Overview}

\begin{figure*}[!t]
\centering
\includegraphics[width=0.95\textwidth]{"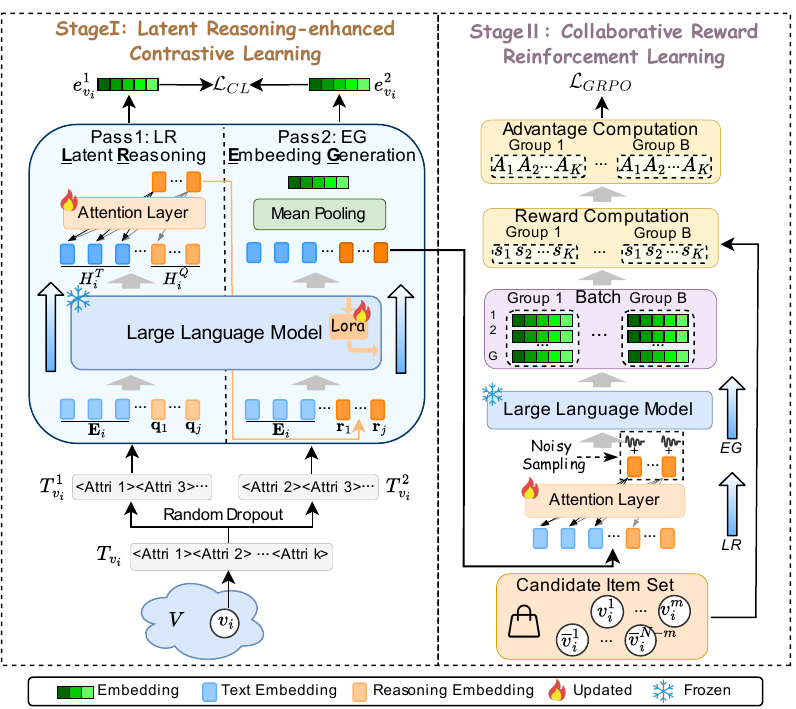"}
\caption{Overview of the proposed \name framework.}
\label{fig:framework}
\vspace{-2mm}
\end{figure*}
 
Figure~\ref{fig:framework} illustrates the overall workflow of the proposed \name, which aims to first enhance LLMs-derived embeddings with deep semantic reasoning and then inject collaborative signals for recommendation. Accordingly, \name is organized as a two-stage training pipeline. The first stage, termed \textbf{Latent Reasoning-enhanced Contrastive Learning} (LRCL), exploits the inner reasoning capacity of LLMs to capture richer item semantics before embedding generation. 
Specifically, LRCL follows a two-forward-pass process: an attention module first generates reasoning tokens, which are then combined with the original text tokens to produce semantically enriched item embeddings. Subsequently, contrastive learning is applied to further learn item semantic information.
The second stage, termed \textbf{Collaborative Reward Reinforcement Learning} (CRRL), aims to explicitly inject collaborative signals into the embedding generator. CRRL treats item co-occurrence as reward feedback and optimizes the lightweight reasoning module with a GRPO-style objective~\cite{shao2024deepseekmath}. 
The reward optimization encourages embeddings of highly co-occurring item pairs to be closer  in embedding space, thereby explicitly reflecting collaborative item-item relations.

\subsection{Latent Reasoning-enhanced Contrastive Learning (LRCL)}

This stage aims to improve the semantic quality of LLMs-based item embeddings. 
Specifically, it first introduces a \textbf{Reasoning-enhanced Embedding Generation} module, which employs an additional attention layer to perform implicit reasoning over the original item text before embedding generation. Then, it employs \textbf{Attribute-level Contrastive Learning} to further train the LLMs into an embedding model that captures the reasoning-enhanced item semantics. Through this stage, the LLMs is equipped with deep semantic embedding capability for item representation.

\subsubsection{\textbf{Reasoning-enhanced Embedding Generation}}

This module obtains reasoning-enhanced item embeddings through two LLMs forward passes.
The first pass, called \textbf{Latent Reasoning}, forwards the item text through the LLMs and employs a lightweight attention mechanism to generate latent reasoning vectors.
The second pass, called \textbf{Embedding Generation}, concatenates these reasoning vectors with the original input embeddings and forwards them through the LLMs again to produce the final item embedding.

\noindent\textbf{Latent Reasoning}.
For each item textual description $T_i$, we first map its tokens into a token embedding sequence $\mathbf{E}_{i}$.
Then, we append $k$ special reasoning placeholder embeddings after $\mathbf{E}_{i}$:
\begin{equation}
\label{eq:llm_input}
\bar{\mathbf{E}}_{i}=[\mathbf{E}_{i};\mathbf{q}_1,\ldots,\mathbf{q}_k],
\end{equation}
where $\{\mathbf{q}_j\}_{j=1}^{k}$ denote reasoning placeholder embeddings and $k$ represents the steps of the latent reasoning process.
Feeding $\bar{\mathbf{E}}_{i}$ into the frozen LLMs yields the last-layer hidden states $H_{i}$. 
The $H_{i}$ can be partitioned into original textual section, denoted as $H_{i}^{T}=[\mathbf{h}_{i,1},\ldots,\mathbf{h}_{i,n_i}]$, and reasoning section, \ie $H_{i}^{Q}=[\mathbf{h}_{i,n_i+1},\ldots,\mathbf{h}_{i,n_i+k}]$.
Since the reasoning process should be linked to each input token, we introduce a lightweight attention module to transform them into a sequence of latent reasoning tokens. 
Specifically, the $j$-th latent reasoning vector is iteratively generated by attending to both input text and generated reasoning vectors:
\begin{equation}
\label{eq:latent_reasoning_attention}
\mathbf{r}_{i}^{j}
={\delta}
\!\left(
\frac{(\mathbf{h}_{i,n_i+j}\mathbf{W}^{Q})
(\mathbf{C}_{i}^{j}\mathbf{W}^{K})^{\top}}
{\sqrt{d}}
\right)
(\mathbf{C}_{i}^{j}\mathbf{W}^{V}).
\end{equation}
where  $\delta(\cdot)$ denotes the softmax function,  $\mathbf{C}_{i}^{j}=[H_{i}^{T};\mathbf{r}_{i}^{1};\ldots;\mathbf{r}_{i}^{j-1}]$ denotes the context available at the $j$-th reasoning step, with $\mathbf{C}_{i}^{1}=H_{i}^{T}$. $\mathbf{W}^{Q}$, $\mathbf{W}^{K}$, and $\mathbf{W}^{V}$ are learnable parameters of the reasoning module, and $\mathbf{r}_{i}^{j}$ is the $j$-th reasoning vector. 

\noindent\textbf{Embedding Generation}.
After obtaining the latent reasoning vectors $\mathbf{R}_{i}=[\mathbf{r}_{i}^{1},\mathbf{r}_{i}^{2}, \ldots, \mathbf{r}_{i}^{k}]$, we replace the reasoning placeholder embeddings in the Eq.~\eqref{eq:llm_input} with these reasoning vectors, and feed it into the frozen LLMs.
The final item embedding is obtained by applying mean pooling over the output hidden states:
\begin{equation}
\label{eq:reasoning_embedding}
\mathbf{e}_{i}
=
\mathrm{MeanPool}\left(
\mathrm{LLMs}\left([\mathbf{E}_{i};\mathbf{R}_{i}]\right)
\right).
\end{equation}
Through this process, a reasoning-enhanced item embedding $\mathbf{e}_{i}$ is obtained under the guidance of latent reasoning representations.

\subsubsection{\textbf{Attribute-level Contrastive Learning}}

To further facilitate the LLMs as an embedding model for recommendation, LRCL applies attribute-level contrastive learning. For each item $v_i$, we randomly drop part of its textual attributes twice and obtain two augmented descriptions $T_{i}^{1}$ and $T_{i}^{2}$. Following the reasoning-enhanced embedding process described above, we obtain their corresponding embeddings $\mathbf{e}_{i}^{1}$ and $\mathbf{e}_{i}^{2}$.
After that, the contrastive loss for one-side augmentation is defined as:
\begin{equation}
\label{eq:lrcl_loss_one_side}
\mathcal{L}_{\mathrm{CL}}^{1}
=
-\frac{1}{B}
\sum_{i=1}^{B}
\log
\frac{
\exp\left(\mathrm{sim}(\mathbf{e}_{i}^{1},\mathbf{e}_{i}^{2})/\tau\right)
}{
\sum_{k=1}^{B}
\exp\left(\mathrm{sim}(\mathbf{e}_{i}^{1},\mathbf{e}_{k}^{2})/\tau\right)
},
\end{equation}
where $B$ is the batch size, $\mathrm{sim}(\cdot,\cdot)$ denotes the similarity function, and $\tau$ is the temperature coefficient. Similarly, the other-side contrastive loss $\mathcal{L}_{\mathrm{CL}}^{2}$ can be obtained by exchanging the positions of $\mathbf{e}^{1}$ and $\mathbf{e}^{2}$ in Eq.~\eqref{eq:lrcl_loss_one_side}. The final LRCL objective is:
\begin{equation}
\label{eq:lrcl_loss}
\mathcal{L}_{\mathrm{LRCL}}
=
\mathcal{L}_{\mathrm{CL}}^{1}
+
\mathcal{L}_{\mathrm{CL}}^{2}.
\end{equation}

\subsection{Collaborative Reward Reinforcement Learning (CRRL)}

Although LRCL improves the semantic quality of item embeddings, recommendation tasks also depend on collaborative relations among items~\cite{he2020lightgcn}. Therefore, this stage aims to inject item co-occurrence signals into the embedding generator through reward-based optimization. Specifically, it first introduces a \textbf{Co-occurrence Reward Function}. It encourages items with high co-occurrence frequencies with the target item to be closer in the embedding space, thereby explicitly incorporating collaborative signals. Then, it employs \textbf{Batch Relative Policy Optimization} to optimize the lightweight reasoning module with batch-normalized advantages, enableing the generated embeddings to better reflect collaborative item-item relations. 

\subsubsection{\textbf{Co-occurrence Reward Function}}

To construct positive and negative candidates for collaborative signal extraction, we first build item-item co-occurrence statistics from user interaction sequences. Let $c_{ij}$ denote the times of item $v_i$ and item $v_j$ co-occur in user histories. For each anchor item $v_i$, we select the top-$M$ items with the highest co-occurrence counts as its positive items:
\begin{equation}
\mathcal{P}_{i}
=
\operatorname{Top\_M}_{v_j\in\mathcal{V},\,j\neq i}(c_{ij}),
\end{equation}
We then sample $N-M$ negative items from the set of items that have not co-occurred with $v_i$, denoted as $\mathcal{N}_{i}$. Next, the candidate set  can be constructed , \ie $\mathcal{C}_{i}=\mathcal{P}_{i}\cup\mathcal{N}_{i}$, where $|\mathcal{C}_{i}|=N$.

Unlike standard reinforcement learning in NLP tasks, sampling the discrete token, our reasoning vectors lie in a continuous space. To obtain a group of samples for $v_i$, we add Gaussian perturbations to latent reasoning vectors $\mathbf{R}_i$ with $G$ times to simulate different samples:
\begin{equation}
\label{eq:noise_reasoning}
\hat{\mathbf{R}}_{i}^{g}
=
\mathbf{R}_{i}
+
\boldsymbol{\epsilon}^{g},
\end{equation}
where $\boldsymbol{\epsilon}^{g}\sim\mathcal{N}(\mathbf{0},\sigma^{2}), g\in\{1,2,\ldots,G\}$ and $\boldsymbol{\epsilon}^{1}=\mathbf{0}$ is used to keep the original noise-free sample, while $\sigma$ controls the noise strength. Replacing ${\mathbf{R}}_{i}$ with $\hat{\mathbf{R}}_{i}^{g}$ in Eq.~\eqref{eq:reasoning_embedding}, we obtain $G$ sampled embeddings $\{\hat{\mathbf{e}}_{i}^{g}\}_{g=1}^{G}$ for item $v_i$.

The reward aim to encourage the sampled embedding $\hat{\mathbf{e}}_{i}^{g}$ to be closer to highly co-occurring positive items to rank higher within the candidate set. Specifically, each positive item is weighted by its normalized co-occurrence frequency, and the reward is defined as:
\begin{equation}
\label{eq:co_reward}
\begin{aligned}
rank_{i}^{p}
&=
\sum_{v_c\in\mathcal{C}_{i}}
\mathrm{I}\!\left[
\mathrm{sim}(\hat{\mathbf{e}}_{i}^{g},\mathbf{e}_{c})
>
\mathrm{sim}(\hat{\mathbf{e}}_{i}^{g},\mathbf{e}_{p})
\right], \\
s_{i}^{g}
&=
\sum_{v_p\in\mathcal{P}_{i}}
w_{i,p}
\log\left(
1+\frac{1}{1+rank_{i}^{p}}
\right).
\end{aligned}
\end{equation}
where $\mathrm{I}[\cdot]$ is the indicator function, $w_{i,p}={c_{ip}}/{\sum_{v_q\in\mathcal{P}_{i}}c_{iq}}$ is the co-occurrence weight of positive item $v_p$, $\mathbf{e}_{c}$ is the embedding of candidate item $v_c$, and $\mathbf{e}_{p}$ is the embedding of positive item $v_p$.

\subsubsection{\textbf{Batch Relative Policy Optimization}}
After obtaining reward scores, CRRL computes advantages for reinforcement optimization. 
In standard GRPO, rewards are normalized within each group. However, our group samples are generated by perturbing the same latent reasoning tokens. Thus, when the original sample is of low quality, group-wise normalization may amplify unreliable signals. To improve stability, we adopt a batch-normalized advantage by using the noise-free samples as the batch-level reference. Specifically, the batch baseline is computed as $\bar{s}_{\mathcal{B}}=\frac{1}{|\mathcal{V}_{\mathcal{B}}|}\sum_{v_i\in\mathcal{V}_{\mathcal{B}}}s_i^1$, where $\mathcal{V}_{\mathcal{B}}$ denotes the item set in the current mini-batch. The advantage of the $g$-th sampled embedding of item $v_i$ is:
\begin{equation}
\label{eq:batch_advantage}
A_i^{g}
=
\frac{
s_i^{g}-\bar{s}_{\mathcal{B}}
}{
\left\|
\mathbf{S}_{\mathcal{B}}-\bar{s}_{\mathcal{B}}
\right\|_2
+
\epsilon
},
\end{equation}
where $\mathbf{S}_{\mathcal{B}}$ denotes the batch reward scores,  $\|\cdot\|_2$ denotes L2 normalization and $\epsilon$ is a small constant for numerical stability.

With the computed advantage, CRRL updates the lightweight reasoning module using a GRPO-style clipped objective. 
The importance ratio is,
\begin{equation}
\label{eq:importance_ratio}
\eta_i^g(\phi)
=
\frac{
\pi_{\phi}(\hat{\mathbf{e}}_{i}^{g}\mid T_{v_i},\hat{\mathbf{r}}_{i}^{g})
}{
\pi_{\phi_{\mathrm{old}}}(\hat{\mathbf{e}}_{i}^{g}\mid T_{v_i},\hat{\mathbf{r}}_{i}^{g})
}.
\end{equation}
As a result, the CRRL objective is,
\begin{equation}
\label{eq:grpo_objective}
\begin{aligned}
\mathcal{L}_{\mathrm{CRRL}}
=
-\sum_{v_i\in\mathcal{V}_{\mathcal{B}} }
\frac{1}{G}
\sum_{g=1}^{G}
\min\Big(
&\eta_i^g A_i^g,  \\
\mathrm{clip}(\eta_i^g,1-\epsilon_c,1+\epsilon_c)A_i^g
\Big) 
&+\beta D_{\mathrm{KL}}(\pi_{\phi}\Vert\pi_{\mathrm{ref}}).
\end{aligned}
\end{equation}
where $\epsilon_c$ is the clipping coefficient, and $\beta$ controls the KL regularization strength. Since the base LLMs is frozen and only the lightweight reasoning module is updated, the KL regularization term $D_{KL}$ evaluates to zero in practice.

\subsection{Training and Inference}

In this section, we detail the training and inference process of \name.

\noindent\textbf{Training.}
During the LRCL stage, we adopt LoRA~\cite{hu2022lora} to fine-tune the LLMs in a parameter-efficient manner. Let 
$\Lambda=\{\mathbf{A}_{\ell},\mathbf{B}_{\ell}\}_{\ell=1}^{L}$ denote the trainable low-rank LoRA matrices inserted into $L$ transformer layers in LLMs, and $\Phi=\{\mathbf{W}^{Q},\mathbf{W}^{K},\mathbf{W}^{V}\}$ denote the parameters of the lightweight reasoning attention module. 
The LRCL objective is:
\begin{equation}
\label{eq:LRCL_objective}
\arg\min_{\Lambda,\Phi}\mathcal{L}_{\mathrm{LRCL}}.
\end{equation}
After LRCL converges, we enter the CRRL stage. In this stage, 
the LoRA parameters $\Lambda$ learned in the first stage are fixed, and only the reasoning attention module $\Phi$ is updated to preserve the semantic representation ability. Thus, the objective is:
\begin{equation}
\label{eq:CRRL_objective}
\arg\min_{\Phi}\mathcal{L}_{\mathrm{CRRL}}.
\end{equation}
\vspace{-2pt}
After the embedding generator is trained, we precompute the item embeddings $\{\mathbf{e}_{v_i}\}_{v_i\in\mathcal{V}}$ and use them to initialize the item embedding layer of the downstream SRS backbone. 
The SRS backbone is then trained with pairwise binary cross-entropy loss for next-item prediction.

\noindent\textbf{Inference.}
During inference, \name follows the standard \textit{Embedding} and \textit{Sequence} pipeline of SRS. In the \textit{Embedding} step, all item embeddings are generated offline and cached in advance, then loaded as the embedding layer of the SRS backbone. In the \textit{Sequence} step, the backbone consumes the cached embedding sequence of a user's history and outputs the next-item prediction. Therefore, \name requires no online LLMs computation and introduces no additional inference burden compared to traditional SRS models.

\begin{algorithm}[!t]
\caption{Training and inference process of \name}
\label{alg:think2rec}
\raggedright
\small
\begin{algorithmic}[1]
\Statex \textbf{Initialization}
\State Initialize the LoRA parameters $\Lambda$, reasoning module $\Phi$, and SRS backbone $f_{\Theta}$.
\State Construct the co-occurrence statistics $\{c_{ij}\}$ from user sequences.
\Statex \textbf{Training Process}
\Statex \textit{LRCL stage}
\State Build two attribute-dropout views for each item.
\State Generate latent reasoning tokens and reasoning-enhanced embeddings by Eq.~\eqref{eq:latent_reasoning_attention} and Eq.~\eqref{eq:reasoning_embedding}.
\State Optimize $\mathcal{L}_{\mathrm{LRCL}}$ to update $\Lambda$ and $\Phi$.
\Statex \textit{CRRL stage}
\State  Perform Latent Reasoning to get latent reasoning tokens of each items. 
\State Perturb latent reasoning tokens to obtain group embeddings.
\State Compute rewards and advantages by Eq.~\eqref{eq:co_reward} and Eq.~\eqref{eq:batch_advantage}.
\State Optimize $\mathcal{L}_{\mathrm{CRRL}}$ to update $\Phi$.
\Statex \textit{SRS training stage}
\State Precompute item embeddings $\{\mathbf{e}_{v_i}\}_{v_i\in\mathcal{V}}$ through the well-trained LLM and cache them for downstream SRS training.   
\State Load them as the embedding layer of $f_{\Theta}$, and train the SRS backbone with its original recommendation objective.
\Statex \textbf{Inference Process}
\State Initial embedding layer of  $f_{\Theta}$ with the cached item embedding.
\State Feed users' historical item embedding sequences into $f_{\Theta}$to produce next-item prediction. 
\end{algorithmic}
\end{algorithm}

\section{Experiment}

In this section, we present the experimental results to answer the following Research Questions (RQ).
\begin{itemize}[leftmargin=*, itemsep=2pt, topsep=0pt, parsep=0pt, partopsep=0pt]
    \item \textbf{RQ1}: How does the proposed \name perform compared with existing LLMs-based embedding methods? Can \name enhance different SRS backbones?
    \item \textbf{RQ2}: Do the key designs of \name take effect?
    \item \textbf{RQ3}: How do important hyper-parameters affect the performance of \name?
    \item \textbf{RQ4}: Can \name alleviate the long-tail problem while maintaining competitive performance on popular items?
    \item \textbf{RQ5}: Can \name maintain stable effectiveness across different LLM series?
\end{itemize}

\subsection{Experimental Settings}

\begin{table}[!t]
\centering
\tabcolsep=0.1cm
\caption{Statistics of the three benchmark datasets.}
\resizebox{1\linewidth}{!}{
\begin{tabular}{lccccc}
\toprule
\textbf{Dataset} & \textbf{\#Users} & \textbf{\#Items} & \textbf{\#Inter.} & \textbf{Avg. Len.}& \textbf{Sparsity} \\
\midrule
Yelp & 35,417 & 16,401 & 256,339 & 7.49 & 99.95\% \\
CD& 68,249 & 23,971 & 289,018 & 4.23 & 99.98\% \\
Games& 50,207 & 16,142 & 364,431 & 7.26 & 99.96\% \\
\bottomrule
\end{tabular}
}
\label{tab:dataset_statistics}
\end{table}

\subsubsection{\textbf{Datasets}}
We evaluate the proposed \name framework on three widely used real-world benchmark datasets: Yelp, Amazon CD, and Amazon Games.
Yelp is constructed from business reviews in the Yelp Open Dataset\footnote{\url{https://www.yelp.com/dataset}}.
Amazon CD and Amazon Games are two representative categories from the Amazon review dataset\footnote{\url{https://cseweb.ucsd.edu/~jmcauley/datasets.html\#amazon_reviews}}, containing user interactions with music and video game products.

Following prior sequential recommendation studies~\cite{kang2018self,sun2019bert4rec}, each dataset is preprocessed and organized into user interaction sequences according to timestamps.
For preprocessing, we apply dataset-specific filtering strategies to obtain reliable user behavior sequences and avoid the cold-start issue.
For the Yelp and Amazon Games datasets, users with fewer than $5$ interactions and items with fewer than $3$ interaction records are removed.
For the Amazon CD dataset, users and items with fewer than $3$ interactions are filtered out.
For training and evaluation, we adopt the leave-one-out protocol in sequential recommendation.
The statistics of the processed datasets are summarized in Table~\ref{tab:dataset_statistics}.

\subsubsection{\textbf{Baselines}}
To evaluate the effectiveness of the proposed \name framework, we compare it with the following representative LLMs-based embedding methods for sequential recommendation:
\begin{itemize}[leftmargin=*, itemsep=2pt, topsep=0pt, parsep=0pt, partopsep=0pt]
    \item \textbf{LLM2Vec} ~\cite{liu2025llmemb} adapts decoder-only LLMs into effective text embedding models through bidirectional attention modification and unsupervised contrastive learning.
    \item \textbf{TSLRec} ~\cite{liu2024practice} injects collaborative information into LLMs through user-level preference distribution reconstruction and fuses LLM knowledge-enhanced embeddings with randomly initialized item embeddings via gating.
    \item \textbf{SAID} ~\cite{hu2024enhancing} aligns item ID embeddings with semantic representations generated by LLMs through a lightweight projector module.
    \item \textbf{LLMEmb} ~\cite{liu2025llmemb} adopts attribute-level augmentation and contrastive learning to improve LLM-generated item embeddings, followed by recommendation adaptation training.
    \item \textbf{LLM2Rec} ~\cite{he2025llm2rec} transforms LLMs into recommendation-oriented embedding models through collaborative supervised fine-tuning and item-level embedding optimization.
\end{itemize}

\begin{table*}[!t]
\small
\setlength{\tabcolsep}{2pt}
\caption{Overall and long-tail performance comparison on three datasets. The best results are highlighted in bold,  the second-best results are underlined, and “*” indicates the statistically significant improvements (i.e., two-sided t-test with $p<0.05$) over the best baseline.
  }
\label{tab:main_results}
\resizebox{\textwidth}{!}{
\begin{tabular}{ll|cccc|cccc|cccc}
\toprule
& & \multicolumn{4}{c|}{\textbf{Yelp}}
& \multicolumn{4}{c|}{\textbf{CD}}
& \multicolumn{4}{c}{\textbf{Games}} \\
\cmidrule(lr){3-6}\cmidrule(lr){7-10}\cmidrule(lr){11-14}
\textbf{Backbone}& \textbf{Model}& \multicolumn{2}{c}{\textbf{Overall}} & \multicolumn{2}{c|}{\textbf{Tail}} & \multicolumn{2}{c}{\textbf{Overall}} & \multicolumn{2}{c|}{\textbf{Tail}} & \multicolumn{2}{c}{\textbf{Overall}} & \multicolumn{2}{c}{\textbf{Tail}} \\
\cmidrule(lr){3-4}\cmidrule(lr){5-6}\cmidrule(lr){7-8}\cmidrule(lr){9-10}\cmidrule(lr){11-12}\cmidrule(lr){13-14}
& & H@10 & N@10 & H@10 & N@10 & H@10 & N@10 & H@10 & N@10 & H@10 & N@10 & H@10 & N@10 \\
\midrule
\multirow{6}{*}{SASRec}
& LLM2Vec & 0.7567 & 0.4785 & 0.5806 & 0.3043 & \underline{0.5117} & 0.3143 & \underline{0.4146} & 0.2353 & 0.5355 & 0.3265 & 0.3520 & 0.1928 \\
& TSLRec & \underline{0.7608} & \underline{0.4969} & 0.5151 & 0.2494 & 0.5103 & \underline{0.3529} & 0.2339 & 0.1336 & 0.4923 & 0.2992 & 0.0072 & 0.0026 \\
& SAID & 0.6744 & 0.3825 & \underline{0.6157} & \underline{0.3120} & 0.4842 & 0.3280 & 0.3369 & 0.2076 & 0.4400 & 0.2594 & 0.3018 & 0.1523 \\
& LLMEmb & 0.6752 & 0.4259 & 0.4787 & 0.2619 & 0.5098 & 0.3215 & 0.4048 & \underline{0.2353} & \underline{0.5583} & \underline{0.3493} & \underline{0.3981} & \underline{0.2229} \\
& LLM2Rec & 0.3068 & 0.1706 & 0.1675 & 0.0897 & 0.4350 & 0.2602 & 0.2704 & 0.1453 & 0.5498 & 0.3306 & 0.2920 & 0.1452 \\
&\cellcolor{orange!10}\textbf{\name} & \cellcolor{orange!10}\textbf{0.7862\textsuperscript{*}} & \cellcolor{orange!10}\textbf{0.5036\textsuperscript{*}} & \cellcolor{orange!10}\textbf{0.6504\textsuperscript{*}} & \cellcolor{orange!10}\textbf{0.3446\textsuperscript{*}} & \cellcolor{orange!10}\textbf{0.5581\textsuperscript{*}} & \cellcolor{orange!10}\textbf{0.3637\textsuperscript{*}} & \cellcolor{orange!10}\textbf{0.4599\textsuperscript{*}} & \cellcolor{orange!10}\textbf{0.2778\textsuperscript{*}} & \cellcolor{orange!10}\textbf{0.5638\textsuperscript{*}} & \cellcolor{orange!10}\textbf{0.3557\textsuperscript{*}} & \cellcolor{orange!10}\textbf{0.4022\textsuperscript{*}} & \cellcolor{orange!10}\textbf{0.2284\textsuperscript{*}} \\
\midrule
\multirow{6}{*}{GRU4Rec}
& LLM2Vec & \underline{0.6155} & \underline{0.3805} & \underline{0.3877} & \underline{0.1974} & 0.4324 & 0.2662 & 0.3351 & \underline{0.1890} & 0.5121 & 0.3052 & 0.3418 & 0.1842 \\
& TSLRec & 0.5206 & 0.3381 & 0.1554 & 0.0732 & \underline{0.4718} & \underline{0.3183} & 0.2226 & 0.1228 & 0.4908 & 0.2997 & 0.0525 & 0.0228 \\
& SAID & 0.5124 & 0.2781 & 0.3336 & 0.1708 & 0.2831 & 0.1820 & 0.1511 & 0.0827 & 0.4098 & 0.2336 & 0.2422 & 0.1197 \\
& LLMEmb & 0.6131 & 0.3748 & 0.3581 & 0.1835 & 0.4619 & 0.2640 & \underline{0.3496} & 0.1820 & \underline{0.5201} & \underline{0.3092} & \underline{0.3473} & \underline{0.1843} \\
& LLM2Rec & 0.2308 & 0.1190 & 0.1426 & 0.0697 & 0.3528 & 0.2085 & 0.2216 & 0.1182 & 0.4822 & 0.2848 & 0.2706 & 0.1379 \\
&\cellcolor{orange!10}\textbf{\name} & \cellcolor{orange!10}\textbf{0.7566\textsuperscript{*}} & \cellcolor{orange!10}\textbf{0.4749\textsuperscript{*}} & \cellcolor{orange!10}\textbf{0.5663\textsuperscript{*}} & \cellcolor{orange!10}\textbf{0.2812\textsuperscript{*}} & \cellcolor{orange!10}\textbf{0.5390\textsuperscript{*}} & \cellcolor{orange!10}\textbf{0.3293\textsuperscript{*}} & \cellcolor{orange!10}\textbf{0.4295\textsuperscript{*}} & \cellcolor{orange!10}\textbf{0.2401\textsuperscript{*}} & \cellcolor{orange!10}\textbf{0.5237\textsuperscript{*}} & \cellcolor{orange!10}\textbf{0.3140\textsuperscript{*}} & \cellcolor{orange!10}\textbf{0.3608\textsuperscript{*}} & \cellcolor{orange!10}\textbf{0.1941\textsuperscript{*}} \\
\midrule
\multirow{6}{*}{BERT4Rec}
& LLM2Vec & \underline{0.7011} & 0.4337 & 0.4551 & 0.2119 & 0.4756 & 0.2808 & 0.3563 & 0.1893 & 0.5474 & 0.3323 & 0.3306 & 0.1710 \\
& TSLRec & 0.6819 & \underline{0.4417} & 0.3491 & 0.1465 & 0.5213 & \underline{0.3432} & 0.1837 & 0.0974 & 0.5415 & 0.3408 & 0.0434 & 0.0178 \\
& SAID & 0.2927 & 0.1673 & 0.1447 & 0.0636 & 0.4735 & 0.2965 & 0.3323 & 0.1895 & 0.4218 & 0.2337 & 0.3037 & 0.1560 \\
& LLMEmb & 0.6881 & 0.4173 & \underline{0.4858} & \underline{0.2443} & \underline{0.5312} & 0.3277 & \underline{0.4140} & \underline{0.2291} & \underline{0.5576} & \underline{0.3409} & \underline{0.3728} & \underline{0.1972} \\
& LLM2Rec & 0.2942 & 0.1607 & 0.1324 & 0.0638 & 0.4174 & 0.2399 & 0.2233 & 0.1095 & 0.5383 & 0.3197 & 0.2496 & 0.1188 \\
&\cellcolor{orange!10}\textbf{\name} & \cellcolor{orange!10}\textbf{0.7793\textsuperscript{*}} & \cellcolor{orange!10}\textbf{0.4838\textsuperscript{*}} & \cellcolor{orange!10}\textbf{0.6274\textsuperscript{*}} & \cellcolor{orange!10}\textbf{0.3079\textsuperscript{*}} & \cellcolor{orange!10}\textbf{0.5831\textsuperscript{*}} & \cellcolor{orange!10}\textbf{0.3721\textsuperscript{*}} & \cellcolor{orange!10}\textbf{0.4738\textsuperscript{*}} & \cellcolor{orange!10}\textbf{0.2743\textsuperscript{*}} & \cellcolor{orange!10}\textbf{0.5702\textsuperscript{*}} & \cellcolor{orange!10}\textbf{0.3517\textsuperscript{*}} & \cellcolor{orange!10}\textbf{0.3865\textsuperscript{*}} & \cellcolor{orange!10}\textbf{0.2091\textsuperscript{*}} \\
\bottomrule
\end{tabular}
}
\vspace{-2mm}
\end{table*}

\subsubsection{\textbf{Implementation Details}}
All experiments are conducted on a Linux server equipped with AMD EPYC 7543 32-Core processors and four NVIDIA GeForce RTX 3090 GPU with 24GB memory.
The implementation is based on Python~3.10.20, PyTorch~2.7.1+cu118, and CUDA~12.9.
For LLMs-based embedding generation, we employ Qwen2.5-0.5B as the backbone LLMs to ensure a fair comparison.

\subsubsection{\textbf{Evaluation Metrics}}
We evaluate all models under the standard \textit{Top-10} ranking setting.
Specifically, two commonly used metrics are adopted: \textit{Hit Rate} (\textbf{H@10}) and \textit{Normalized Discounted Cumulative Gain} (\textbf{N@10}).
Following~\cite{kang2018self}, for each user, we randomly sample 100 items that the user has not interacted with as negative candidates, combined with the ground-truth positive item to compute the evaluation metrics.

\subsection{Overall Performance (RQ1)}

To respond to RQ1, we report both overall and tail-item performance of our method and all baselines under three sequential recommendation backbone models on three datasets, as shown in Table~\ref{tab:main_results}.
Specifically, according to the Pareto principle~\cite{newman2005power}, items whose popularity ranks fall into the last 80\% are divided into the tail group.

The results indicate that \name consistently outperforms all competitors in both overall and long-tail recommendation. For a more detailed analysis,
TSLRec achieves competitive overall performance in several settings because its user-level preference reconstruction task injects collaborative information into LLMs.
However, its long-tail performance is much weaker, since the preference reconstruction relies on item collaborative information, which is unfriendly for long-tail items.
In contrast, \name leads overall recommendation while maintaining stronger long-tail performance, better balancing the trade-off between head and tail items.
For the other baselines, SAID performs relatively weaker since it only uses LLMs semantics to train a projector, which may be insufficient to fully capture item semantics.
LLM2Vec and LLMEmb show strong competitiveness, but they still fall behind \name, because they do not explicitly exploit the reasoning capability of LLMs and incorporate collaborative signals into item representations. 
LLM2Rec performs well on overall recommendation but is less efficient on tail items, since injecting collaborative signals through SFT may disturb the original semantic space of the LLMs and thus hurt long-tail item representation.
These results demonstrate that \name can better explore item semantics and effectively incorporate collaborative signals. Furthermore, it can consistently adapt to different SRS backbones.


\subsection{Ablation Study (RQ2)}

\begin{table}[htbp]
\centering
\small
\setlength{\tabcolsep}{5pt}
\caption{The ablation study conducted on the CD dataset based on the SASRec backbone. The best results are highlighted in bold.}
\label{tab:ablation_study}
\resizebox{\columnwidth}{!}{
\begin{tabular}{l|cccc}
\toprule[1.2pt]
\multirow{2}{*}{\textbf{Model}}
& \multicolumn{2}{c}{\textbf{Overall}}
& \multicolumn{2}{c}{\textbf{Tail}} \\
\cmidrule(lr){2-5}
& H@10 & N@10 & H@10 & N@10 \\
\midrule
\cellcolor{orange!10}\textbf{ReaEmb} & \cellcolor{orange!10}\textbf{0.5581} & \cellcolor{orange!10}\textbf{0.3637} & \cellcolor{orange!10}\textbf{0.4599} & \cellcolor{orange!10}\textbf{0.2778} \\
\textit{w/o} Reasoning & 0.5259 & 0.3387 & 0.4378 & 0.2658 \\
\textit{w/o} RL & 0.5484 & 0.3605 & 0.4528 & 0.2765 \\
\textit{w/o} atten & 0.5362 & 0.3466 & 0.4392 & 0.2677 \\
\textit{w/o} BRPO & 0.5490 & 0.3614 & 0.4511 & 0.2756 \\
\bottomrule[1.2pt]
\end{tabular}
}
\end{table}

To respond to RQ2, we conduct the ablation study and show the results in Table~\ref{tab:ablation_study}.
To investigate the effect of LRCL, \textit{w/o} Reasoning eliminates the first-stage contrastive learning, which leads to a significant performance degradation.
This result demonstrates the necessity of adapting the LLMs into a recommendation-oriented embedding model with reasoning-enhanced item semantics.
Then, \textit{w/o} RL removes the second-stage reinforcement learning.
The performance drops indicates the effectiveness of explicitly injecting collaborative signals with the co-occurrence-based reward.
Moreover, \textit{w/o} atten removes the additional attention module in implicit reasoning.
The consistent decrease shows that the proposed attention module can effectively transform thought placeholders into meaningful latent reasoning tokens.
Finally, \textit{w/o} BRPO replaces the proposed batch-level advantage estimation with group-level advantage computation.
The decline suggests that group-level normalization may produce unstable references under noise-injected sampling settings, whereas batch-level advantages offer more reliable reward guidance.

\subsection{Hyper-parameter Analysis (RQ3)}

\begin{figure}[htbp]
    \centering
    \makebox[\columnwidth][c]{%
        \includegraphics[width=1.05\columnwidth]{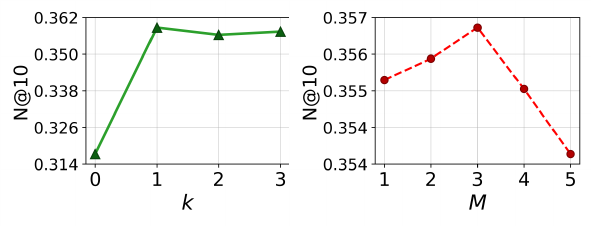}
 }
    \caption{The experimental results for the reasoning step $k$ and the number of positive candidate items $M$ based on the Games dataset and SASRec backbone. }
    \label{fig:hyper_parameter}
\end{figure}

To respond to RQ3, we investigate two important hyper-parameters in \name, i.e., the reasoning step $k$ and the number of positive candidate items $M$. 
We show their trends in Figure~\ref{fig:hyper_parameter}.
As $k$ increases from $0$ to $1$, the H@10 value rises sharply, demonstrating that introducing latent reasoning tokens can effectively capture item semantics beyond direct LLMs encoding.
When $k$ further increases, the H@10 value tends to remain stable.
This indicates that one reasoning token is sufficient to capture most of the reasoning information.
In terms of $M$, the N@10 value first increases as $M$ grows from $1$ to $3$, which demonstrates that more positive collaborative candidates provide richer co-occurrence supervision for reward optimization.
When $M$ further increases, the performance drops, since low-ranked co-occurring items may contain noisy or weak collaborative relations and thus mislead the reinforcement learning process.

\subsection{Group Analysis (RQ4)}

\begin{figure}[!t]
    \centering
    \makebox[\columnwidth][c]{%
        \includegraphics[width=0.95\columnwidth]{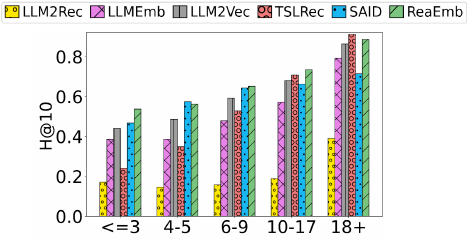}
 }
    \caption{The experimental results of group analysis based on Yelp dataset and SASRec backbone.}
    \label{fig:group_study}
\end{figure}

To respond to RQ4, we further explore the long-tail problem by dividing items into five groups according to their interaction frequency and report the H@10 results in Figure~\ref{fig:group_study}.
Compared with LLM2Rec, LLMEmb and LLM2Vec, \name consistently achieves better performance across all popularity groups, indicating that it can more effectively supplement semantic information and inject collaborative signals.
However, \name underperforms SAID on long-tail groups (i.e., 4--5) and TSLRec on the popular-item group (i.e., 18+).
This is because both methods show a clear seesaw effect: SAID emphasizes semantic information but performs poorly on popular items, while TSLRec relies more on collaborative signals and substantially weakens long-tail recommendation.
In contrast, \name achieves strong performance and better balances these two types of signals.


\subsection{LLMs Series Analysis (RQ5)}

\begin{figure}[!t]
    \centering
    \makebox[\columnwidth][c]{%
        \includegraphics[width=1.05\columnwidth]{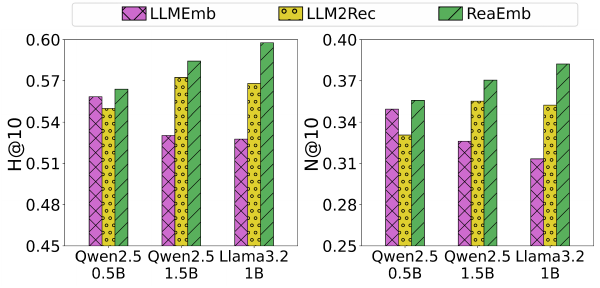}
 }
    \caption{The experimental results of three LLMs based on Games dataset and SASRec backbone.}
    \label{fig:LLM_series}
\end{figure}

To respond to RQ5, we investigate the adaptability and effectiveness of \name across different LLMs from two perspectives: model size and model type, as shown in Figure~\ref{fig:LLM_series}.
For LLMs of different sizes, we compare Qwen2.5-0.5B and Qwen2.5-1.5B.
The results show that \name consistently outperforms LLM2Rec and LLMEmb on both H@10 and N@10, and the larger Qwen2.5-1.5B further improves the performance, indicating that \name can benefit from stronger semantic modeling ability.
For LLMs of different types, we further introduce Llama3.2-1B for comparison.
\name still achieves the best results, while the baselines show some instability, verifying that \name maintains stable superiority across different LLM architectures.

\section{Related Works}

\subsection{\textbf{LLMs as Embedding Generators for Sequential Recommendation}}

To address the long-tail problem in ID-based sequential recommendation~\cite{kang2018self,sun2019bert4rec}, a growing body of research introduces item semantic information to complement collaborative signals.
With the development of LLMs, recent studies have explored how to use their semantic understanding ability to generate item embeddings for sequential recommendation.
One line of work mainly injects semantic information into item representations.
LLM2Vec~\cite{behnamghader2024llm2vec} converts decoder-only LLMs into text encoders through bidirectional attention and contrastive learning; LLM2X~\cite{harte2023leveraging} initializes recommender item embeddings with LLM-generated representations; SAID~\cite{hu2024enhancing} aligns item IDs with the LLM semantic space through projection; and LLMEmb~\cite{liu2025llmemb} uses attribute augmentation and contrastive tuning to generate item embeddings.
Another line further incorporates collaborative information into LLM-based embedding generation.
LLM2Rec~\cite{he2025llm2rec} implicitly injects collaborative signals by next token prediction on item textual sequences; TSLRec~\cite{liu2024practice} uses user-level preference reconstruction to incorporate collaborative knowledge.
Different from these methods, \name introduces latent reasoning before embedding generation to capture deeper semantics and explicitly injects collaborative signals based on item co-occurrence statistics.

\subsection{\textbf{Reinforcement Learning for LLM-based Recommendation}}
Recent studies have started to apply reinforcement learning to LLM-based recommendation to optimize recommendation feedback beyond supervised imitation.
Most existing methods use RL to train LLMs for direct item prediction.
Wang \etal~\cite{wang2024reinforcement} use LLMs for state, reward, and action modeling in offline RL; Rec-R1~\cite{lin2025recr1} optimizes LLM outputs with rewards from a fixed recommender; RecLLM-R1~\cite{xie2025recllm} combines SFT, GRPO, and chain-of-thought rewards for generative recommendation; and LatentR3~\cite{zhang2025reinforced} optimizes item prediction with a perplexity-based reward.
Despite their effectiveness, they still rely on LLMs to directly generate recommended items, causing high online inference latency.
In contrast, \name uses reinforcement learning to inject collaborative signals into item embeddings, enabling efficient recommendation without online LLM generation.

\section{Conclusion}

In this paper, we propose \name, a two-stage LLM-based embedding generator for sequential recommendation.
We first apply Latent Reasoning-enhanced Contrastive Learning to exploit the inner reasoning capacity of LLMs and strengthen item semantic representations, thus alleviating the insufficient semantic understanding problem.
Further, we introduce Collaborative Reward Reinforcement Learning to optimize the lightweight reasoning module with co-occurrence-based rewards, thereby explicitly injecting collaborative signals.
Through comprehensive experiments on three real-world datasets, we validate the effectiveness and flexibility of \name.


\bibliography{main}

\end{document}